\documentclass[twocolumn]{aastex}%\documentclass[12pt,preprint]{aastex}

\newcommand\aastex{AAS\TeX}

\submitjournal{ApJ}
\shorttitle{\aastex\ Discovery of three new millisecond pulsars in Terzan~5}
\shortauthors{Cadelano et al.}

\usepackage{amsmath}

\begin{document} 

\title{Discovery of three new millisecond pulsars in Terzan 5}

\correspondingauthor{Mario Cadelano}
\email{mario.cadelano@unibo.it}

\author{M. Cadelano}
\affiliation{Dipartimento di Fisica e Astronomia, Via Gobetti 93/2 40129 Bologna - Italy}
\affiliation{INAF - Osservatorio Astronomico di Bologna, Via Gobetti 93/3 40129 Bologna - Italy}

\author{S. M. Ransom}
\affiliation{National Radio Astronomy Observatory, Charlottesville, VA 22903, USA}

\author{P. C. C. Freire}
\affiliation{Max-Planck-Institut f$\ddot{u}$r Radioastronomie MPIfR, Auf dem H$\ddot{u}$gel 69, D-53121 Bonn, Germany}

\author{F. R. Ferraro}
\affiliation{Dipartimento di Fisica e Astronomia, Via Gobetti 93/2 40129 Bologna - Italy}

\author{J.~W.~T.~Hessels}
\affiliation{ASTRON, the Netherlands Institute for Radio Astronomy, Postbus 2, 7990 AA, Dwingeloo, The Neherlands}
\affiliation{Anton Pannekoek Institute for Astronomy, University of Amsterdam, Science Park 904, 1098 XH, Amsterdam, The Netherlands}

\author{B. Lanzoni}
\affiliation{Dipartimento di Fisica e Astronomia, Via Gobetti 93/2 40129 Bologna - Italy}

\author{C. Pallanca}
\affiliation{Dipartimento di Fisica e Astronomia, Via Gobetti 93/2 40129 Bologna - Italy}

\author{I. H. Stairs}
\affiliation{Department of Physics and Astronomy, University of British Columbia, 6224 Agricultural Road, Vancouver, BC V6T 1Z1, Canada}

%\author{ M. Cadelano\altaffilmark{1,2}, S. M. Ransom\altaffilmark{3}, P. C. C. Freire\altaffilmark{4}, 
%F. R. Ferraro\altaffilmark{1}, J.~W.~T.~Hessels\altaffilmark{5,6}, B. Lanzoni\altaffilmark{1}, C. Pallanca\altaffilmark{1}  and I. H. Stairs\altaffilmark{7}}
%\affil{\altaffilmark{1} Dipartimento di Fisica e Astronomia,  Universit\`a di Bologna, Via P. Gobetti 93/2, I-40129 Bologna,  Italy }
%\affil{\altaffilmark{2} INAF - Osservatorio Astronomico di Bologna,  Via P. Gobetti 93/3, I-40129 Bologna, Italy }
%\affil{\altaffilmark{3} National Radio Astronomy Observatory, Charlottesville, VA 22903, USA}
%\affil{\altaffilmark{4} Max-Planck-Institut f$\ddot{u}$r Radioastronomie MPIfR, Auf dem H$\ddot{u}$gel 69, D-53121 Bonn, Germany}
%\affil{\altaffilmark{5} ASTRON, the Netherlands Institute for Radio Astronomy, Postbus 2, 7990 AA, Dwingeloo, The Neherlands}
%\affil{\altaffilmark{6} Anton Pannekoek Institute for Astronomy, University of Amsterdam, Science Park 904, 1098 XH, Amsterdam, The Netherlands}
%\affil{\altaffilmark{7} Department of Physics and Astronomy, University of British Columbia, 6224 Agricultural Road, Vancouver, BC V6T 1Z1, Canada}
\begin{abstract}

We report on the discovery of three new millisecond pulsars (namely J1748$-$2446aj,
J1748$-$2446ak and J1748$-$2446al) in the inner regions of the dense stellar system
Terzan 5. These pulsars have been discovered thanks to a { method, alternative to the classical search routines,} that exploited the large set of
archival observations of Terzan~5 acquired with the Green Bank Telescope over 5 years (from
2010 to 2015). This technique allowed the analysis of stacked power spectra obtained by combining 
$\sim$ 206 hours of observation.
J1748$-$2446aj has a spin period of $\sim2.96$ ms, J1748$-$2446ak of
$\sim1.89$ ms (thus it is the fourth fastest pulsar in the cluster) and J1748$-$2446al
of $\sim5.95$ ms. All the three millisecond pulsars are isolated and currently we have timing solutions only for J1748$-$2446aj and J1748$-$2446ak. For these two
systems, we evaluated the contribution to the measured spin-down rate of the
acceleration due to the cluster potential field, thus estimating the intrinsic
spin-down rates, which are in agreement with those typically measured
for millisecond pulsars in globular clusters. Our results increase to 37 the number
of pulsars known in Terzan~5, which now hosts 25\% of the entire pulsar population
identified, so far, in globular clusters.

\end{abstract} 


\section{INTRODUCTION}\label{intro}

Globular clusters (GCs) are the ideal factories for the formation of millisecond
pulsars (MSPs): old and rapidly spinning neutron stars formed in a binary system
through mass and angular momentum accretion from an evolving companion star
\citep[e.g.][]{alpar82, bhattacharya91}. In fact, the large stellar densities in the cores of
GCs favor dynamical interactions, such as exchange interactions and tidal captures,
which can lead to the formation of a large variety of binary systems
whose evolution generates stellar exotica like blue stragglers \citep[e.g.][]{ferraro09, ferraro12},
low-mass X-ray binaries \citep[e.g.][]{pooley03}, cataclysmic variables \citep[e.g.][]{ivanova06} 
and MSPs  \citep[e.g.][]{ferraro01, ferraro03, lorimer03, ransom05, hessels07, pallanca14,cadelano15}. 
%suitable for recycling an old and exhausted neutron star into a
%MSP. 
As a consequence, {the number of MSPs per unit mass in GCs turns out to be $\sim10^3$ times larger} than in
the Galactic field.  Furthermore, dynamical interactions could be also responsible for the
production of exotic MSP systems, 
%not expected from the canonical evolutionary
%scenario, like double neutron star binaries, partially recycled MSPs and also
%isolated MSPs. For similar mechanisms, GCs are also expected to host holy grails of
%the modern pulsar astronomy 
such as double MSP binaries and black hole
- MSP binaries \citep[e.g.][]{freire04,verbunt14}, thus making GCs - particularly those with the denser cores
- intriguing sites where to perform deep pulsar searches in the future.

To date, 146 MSPs are known in 28 GCs\footnote{see \url{http://www.naic.edu/~pfreire/GCpsr.html}} but probably a very large population of several
thousand MSPs is still to be unveiled \citep{bagchi11,chennamangalam13,turk13,hessels15}. The number of MSPs
identified in GCs per year decreased abruptly after 2011, 
thus showing the limit in
%mostly because of the limited 
performance and sensitivity reached by the current generation of
radio telescopes. However, for more than one decade, GCs have been routinely observed
at radio wavelength in order to obtain long term timing solutions of the identified
MPSs. This resulted in the production of a large archive of tens of observations of
the same region of the sky. The work presented here is aimed at showing the large
possibilities of finding new pulsars by exploiting these huge archival data sets. At odds
with traditional pulsar searches, which are based on the analysis of a single and
long time sequence of data, we present here a method to search for pulsars by
incoherently stacking the power spectra obtained from all the available observations. A
similar procedure has been already successfully implemented {in M15 by \citet{anderson93}}, in Terzan 5 by \citet{sulman05}, leading to the discovery of three isolated MSPs (namely J1748$-$2446af, J1748$-$2446ag and J1748$-$2446ah), and in 47 Tucanae by \citet{pan16}, {leading to the discovery of two additional pulsars in the cluster}. {All these pulsars are isolated. Indeed, due to the Doppler shifts of the spin frequency induced by the pulsar motion in binary systems, this method is only effective for discovering isolated pulsars or long period binary pulsars, the latter being less likely to survive in GCs because of stellar encounters.}
%\footnote{{\bf J1748$-$2446ai is an eccentric binary MSP found with classical search routines. See \citet{prager17}}.}

Among the Galactic GCs, Terzan 5 turned out to be the prolific amazing MSP factory. In
fact, 34 MSPs have been identified so far in this system \citep[][Ransom et al. 2018,
in preparation]{ransom05,hessels06,prager17}, which is about $\sim23\%$ of the total number of MSPs
identified in GCs. Terzan 5 is, indeed, one of the most intriguing stellar systems in
the Galaxy. \citet{ferraro09, ferraro16} found out that this system is probably
not a genuine GC, but
more likely the pristine remnant of a building block of the Galactic bulge, which was
originally much more massive than today \citep{lanzoni10}. 
%Its larger initial mass
%combined with the largest collision rate among all the Galactic GCs \citep{lanzoni10,
%bahra13} could explain such a huge MSP population.

Here we present 
%a new method to search for pulsars in GCs and 
the identification of three new MSPs in Terzan 5. In Section~\ref{analysis} we present 
the dataset and the
stacking procedures that led us to the discovery of the new MSPs. In
Section~\ref{results} we describe the main properties of these new systems and their
timing solutions while in Section~\ref{intrinsic} we constrain some of their physical
parameters. Finally we summarize and draw our conclusions in
Section~\ref{conclusions}.

\section{OBSERVATION AND DATA ANALYSIS}
\label{analysis}

\subsection{Dataset and initial data reduction}
The work presented here has been performed by using 33 archival observations of
Terzan~5 obtained with the 100-m Robert C. Byrd Green Bank Telescope (GBT) from
August 2010 to October 2015. Observations were acquired at 1.5 GHz (L-band) and 2.0 GHz (S-band) using
800 MHz of bandwidth, although radio frequency interference (RFI)  excision
reduced the effective bandwidth to $\sim600$ MHz. Only one observation was obtained
at 820 MHz using 200 MHz of bandwidth.  Observation lengths vary from a minimum of 1.5 hours
to 7.5 hours, the latter typical for the majority of the observations. {Overall we have 1 observation obtained at 820 MHz with observation length of 7.3 hours, 20 observations obtained at 1.5 GHz with a total length of about 135 hours and, finally, 12 observations obtained at 2.0 GHz with a total observation length of about 64 hours.} The total
observation length, resulting from the stack of all these observations, is of about 9
days ($\sim 206$ hours).

{ The data recorded by GUPPI were Full Stokes with 10.24 $\mu s$ sampling and 512 channels, each coherently dedispersed in hardware to a DM of $\rm 238 \ pc \ cm^{-3}$, which is close to the cluster average. The total intensity (i.e. sum of two orthogonal polarizations) was extracted from those data and downsampled to 40.96 $\mu s$ resolution for incoherent dedispersion into 23 DM trials}.

The data have been processed using the PRESTO software suite \citep{ransom02}.
We obtained 22 time series per observation ranging from a DM of
$\rm 233 \ pc \ cm^{-3}$ to $\rm 244 \ pc \ cm^{-3}$ and spaced by $\rm 0.5 \ pc \
cm^{-3}$, plus an additional time series at a control DM of $\rm 100 \ pc \ cm^{-3}$. {The time series have been transformed to the barycenter of the solar system using TEMPO\footnote{\url{http://tempo.sourceforge.net}}} \citep{manchester15}. For each sample of the time series of each observation, we subtracted the mean of all
the channels (i.e. we subtracted the DM=$\rm 0 \ pc \ cm^{-3}$ time series) and excised
some interference by removing samples with values higher than $4\sigma$, where
$\sigma$ is the standard deviation of all the sample values. Since we aim at stacking
together the power spectra of observations of different lengths, we manually added
samples with null values to the time series of shorter length, thus
obtaining time series of length equal to that of the longest one. We then applied a
fast Fourier transform to all the time series { and squared the complex amplitudes} to obtain the power spectra. Finally,
in the power spectra, we ignored all the spectral bins expected to contain the powers
of all the known Terzan 5 pulsars and their harmonics (also accounting for the shifts due to the binary pulsar orbital motions). We also excised  the most relevant RFI, {identified by visual inspection of the power spectra}.

%\vspace{0.3cm}
\subsection{Stacking search procedures}
First of all, we normalized all the available power spectra dividing the spectral powers by the local median value. Then, we summed the 33 individual daily power spectra into a stacked power spectrum for each of the 22 DM trials {and the control DM}. { These final stacked power spectra are nearly chi-squared distributed with 66 degrees of freedom.} In
all these stacked spectra, we performed an harmonic sum, by summing to each power bin the powers of
the corresponding harmonic, { from the second up to the eighth}. This has been done to further enhance the
spectral powers of the still unidentified pulsars. At the end of this, we had 22 stacked power
spectra, plus the one at the control DM.  In order to further remove periodic interference that can be
still persistent in high DM spectra, we subtracted from each stacked power spectrum the control
stacked spectrum obtained at DM=$\rm 100 \ pc \ cm^{-3}$. In this way, a large fraction of RFI,
present in both the control and the science power spectra, are removed, leaving the signal of the
still undiscovered pulsars virtually unaffected. 

\begin{figure}
\centering
\includegraphics[width=9cm]{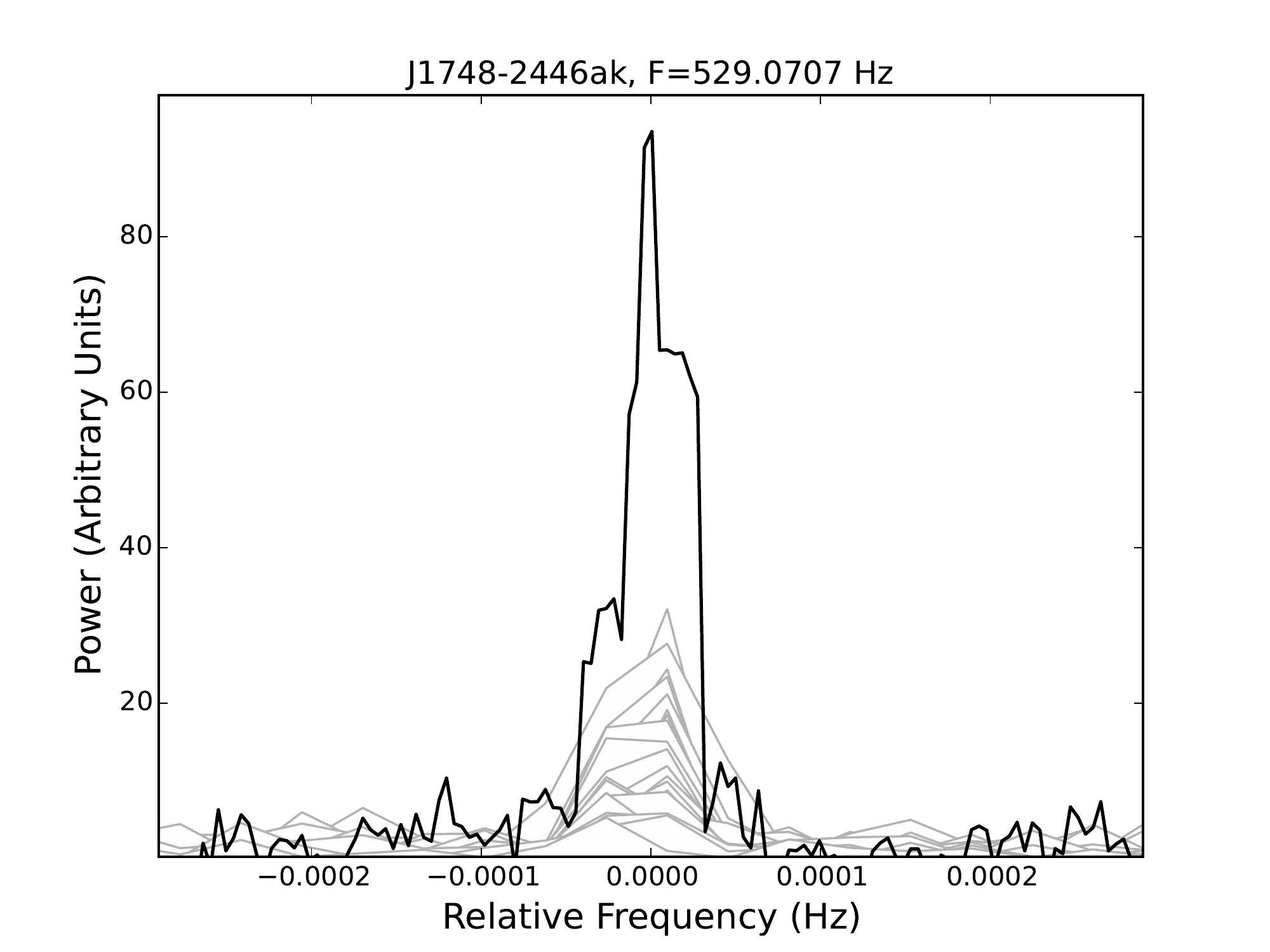}
\caption{Power spectrum around the region of the newly discovered MSP J1748$-$2446ak (Ter5ak). We plotted in black the stacked { and harmonic summed} power spectrum obtained from $\sim215$ hours of archival observations, while in shaded gray we plotted the power spectra obtained from the single daily observations. {The power in the stacked spectrum is spread over more bins than for single observations due to the effect of the harmonic sum}.}
\label{spec}
\end{figure}

In order to illustrate the effectiveness of the method, in Figure~\ref{spec} we compare
 the stacked power spectrum for one of the newly discovered MSPs, with that
obtained from a single observation where this MSP has a high
signal to noise ratio. It can be clearly seen how the stacking procedure greatly enhances the
spectral powers of such a faint object.

%In Figure~\ref{spec} we plot in black the stacked
%power spectrum around the position of one of the newly discovered MSPs, while in red we plot the
%power spectrum, around the same region, obtained from a single observation where this MSP has a high
%signal to noise ratio. It can be clearly seen how the stacking procedure highly enhanced the
%spectral powers of such a faint object.

\begin{figure}[t]
\centering
\includegraphics[width=9cm]{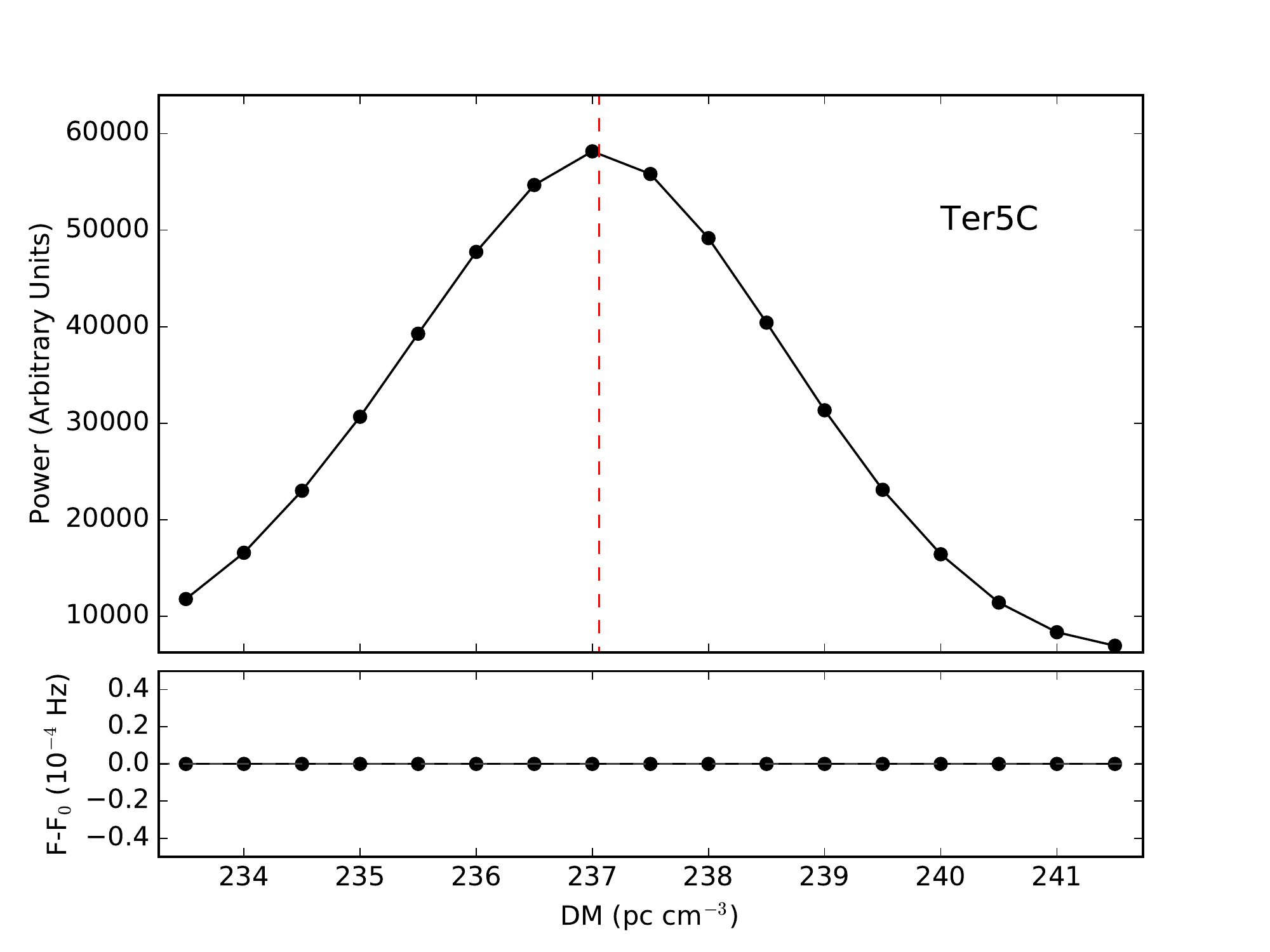}
\caption{{\it Top panel:} Spectral powers of a candidate pulsar, corresponding to J1748$-$2446C (Ter5C), obtained from the stacked power spectra. The powers are plotted as a function of the DM as obtained from the KD-tree algorithm (see text). The red dashed vertical line is the MSP true DM as derived from its timing solution (Ransom et al. 2018, in preparation). {\it Bottom panel:} spin frequency difference across the different DMs in which the candidate has been found. In this case, the candidate has the same frequency in all the DMs, as expected from a real pulsar.}
\label{psrC}
\end{figure}

\begin{figure}
\centering
\includegraphics[width=9cm]{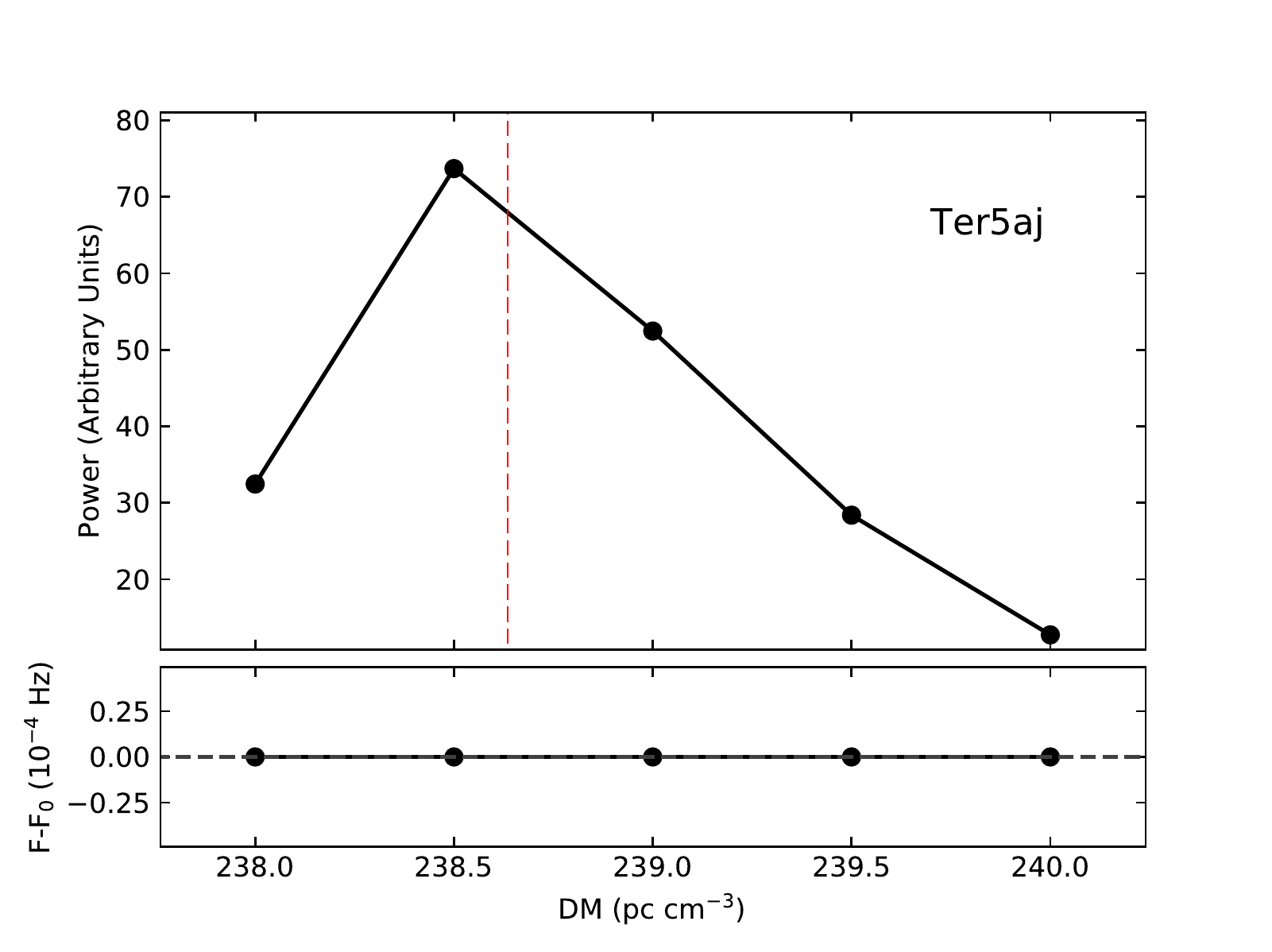}
\caption{Same as in Figure~\ref{psrC} but for the newly discovered pulsar Ter5aj.}
\label{psr_aj}
\end{figure}

The next step is to select, in the stacked power spectra, the periodic signals that likely originate
from a real pulsar instead of from RFI. To do this, we selected in the power spectra all the peaks
above $4.5\sigma$ ($\sigma$ is the standard deviation of the spectral powers) and saved them into a
candidate file. We ended up with 22 candidates files, one per DM, each one containing $\sim13000$
candidate. Therefore the total number of candidates is about $3\times10^{5}$.  We then applied a
KD-tree algorithm to perform a selection of all these candidates. {KD-trees are space-partitioning data structures 
used to organize points in multidimensional spaces and thus useful to perform pulsar searches in these spaces by implementing optimization problems such as the nearest neighbor search}. In fact, in the DM-frequency phase
space, pulsar candidates are expected to be closely segregated around their DM and spin frequency.
On the other hand, RFI can appear in a large range of DMs, also with slightly different spin frequencies.
We used the KD-tree routine included in the Scipy
package\footnote{\url{https://docs.scipy.org/doc/scipy-0.18.1/reference/generated/scipy.spatial.KDTree.html\#scipy.spatial.KDTree}}.
Briefly, we built a 2D tree in a DM-frequency space. Then, for each candidate, we searched for the
1000 closest neighbors and selected only those with a spin frequency compatible (within a tolerance
of $10^{-5}$ Hz) with that of the candidate. Since the pulsar signal is expected to be observed at
contiguous DM values with an almost Gaussian distribution of powers around the true DM value, we
selected as good candidates only those whose closest neighbors are found at contiguous DMs and with
a maximum in the spectral power vs DM space. As an example, we show in Figure~\ref{psrC} the output
of the KD-tree procedure for the case of the known MSP J1748$-$2446C (Ter5C). As can be seen, the
algorithm found, for a candidate corresponding to Ter5C, neighbors at contiguous DMs, with a peak in
the spectral powers very close to the MSP's true DM value. Moreover, the spin frequency does not show
any variation at different DMs, as expected from a real pulsar. The same plot for the newly discovered pulsar Ter5aj is presented in Figure~\ref{psr_aj}. Applying this procedure to our
candidates, { we discarded $\sim97\%$ of them} and ended up with only $\sim100$ possible pulsar
candidates, {most of them close to the chosen threshold of $4.5\sigma$}. These have been individually analyzed, folding the single observations at the candidate
frequency and DM corresponding to that of the maximum spectral power, allowing also a search in spin
period, spin period derivative and DM in order to maximize the signal to noise ratio. {The vast majority of the candidates turned out to be residual RFI, usually very bright only in few observations. Other candidates did not revealed any clear broadband pulsated signal across the observations and thus they have been discarded. We considered as genuine pulsars only those candidates that revealed a clear pulsated signal in more than one observation.}

\section{RESULTS}
\label{results}

%{\bf The minimum detectable flux density reached in a pulsar search can be roughly evaluated using the radiometer equation \citep[see appendix A1.4 of][]{lorimer04}. For a typical 7.5 hours L-band observation of Terzan 5, with an equivalent temperature of the observing system and sky of $T_{sys}\simeq30$ K, considering a 3 ms pulsar with a 10\% duty cycle and a minimum signal to noise ratio of 5, the limiting flux density is of $\sim5 \, \mu Jy$. Stacking together $\sim215$ hours of observations, we reached a limiting flux density of $\sim 1 \, \mu Jy$.}

\begin{figure*}
\centering
\includegraphics[width=12cm]{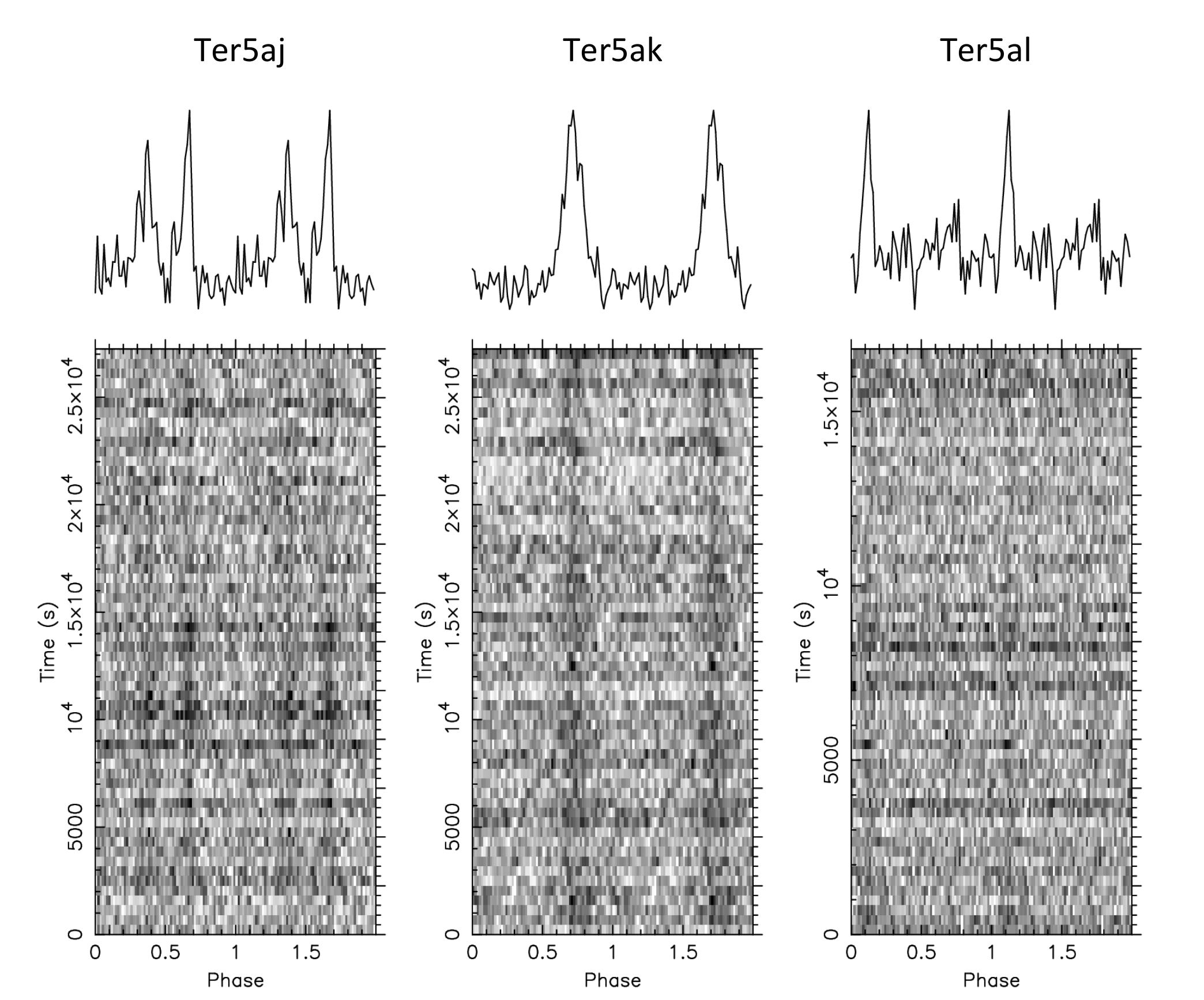}
\caption{{\it Top panels:} Averaged pulse profiles of the best detections of Ter5aj (on the left), Ter5ak (in the middle) and Ter5al (on the right). {\it Bottom panel:} Intensity of the signal (gray scale) as a function of the rotational phase and time for each MSP.}
\label{profili}
\end{figure*}

The method described in the previous section allowed us to discover three previously unknown MSPs in
Terzan~5: J1748$-$2446aj, J1748$-$2446ak and J1748$-$2446al (hereafter, Ter5aj, Ter5ak and
Ter5al, respectively). We plot in Figure~\ref{profili} the average pulse profiles and the signals as
a function of time for these three new MSPs {in the brightest individual days. Moreover, we have been able to blindly re-detect all the other isolated MSPs known in the cluster\footnote{{To do this and to obtain the top panel of Figure~\ref{psrC}, we re run the whole procedure without excising the signal of the known pulsars.}}}.

\subsection{Ter5aj}

\begin{figure}
\centering
\includegraphics[width=9cm]{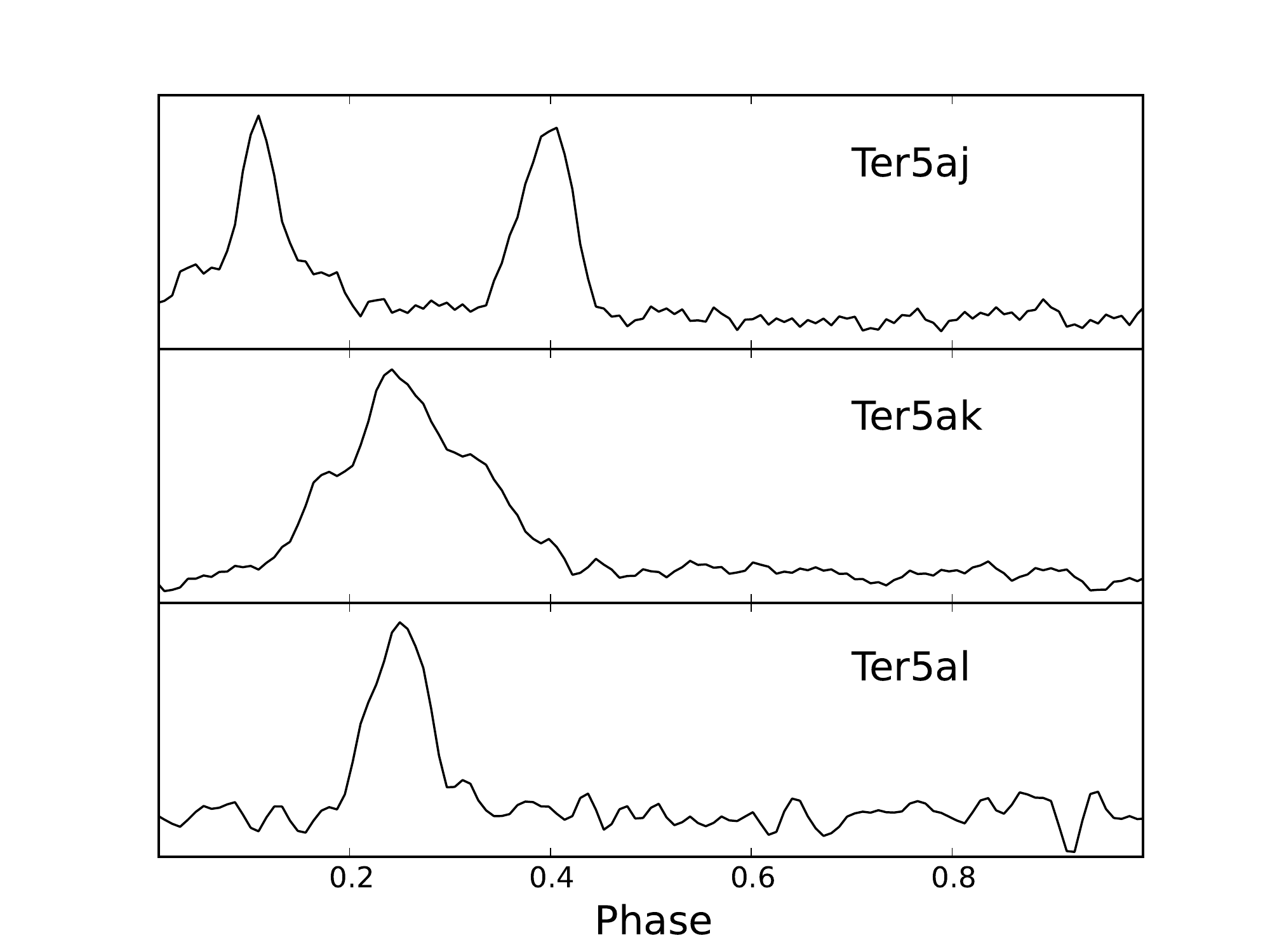}
\caption{Averaged pulse profile of Ter5aj (top panel), Ter5ak (middle panel) and Ter5al (bottom panel), obtained by coherently summing all the GUPPI detections of the MSPs obtained {at a central frequency of 1.4 GHz. The total integration time is of about 135 hours}.}
\label{sumprof}
\end{figure}

\begin{figure}[h]
\centering
\includegraphics[width=9.1cm]{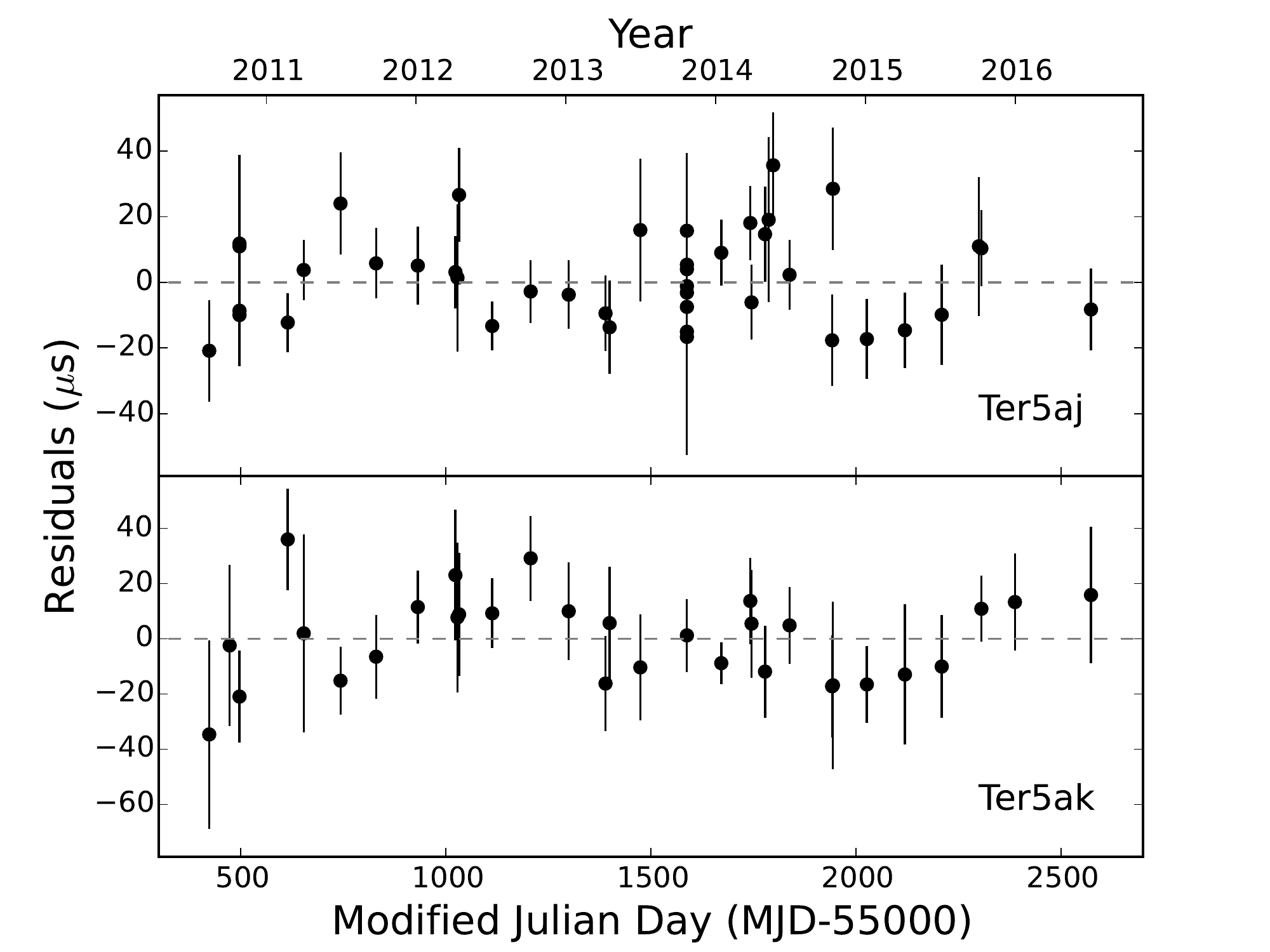}
\caption{Timing residuals for Ter5aj (top panel) and Ter5ak (bottom panel).}
\label{resid}
\end{figure}

Ter5aj has been discovered with a maximum spectral power at $\rm DM=238.50 \ pc \
cm^{-3}$ (see Figure~\ref{psr_aj}). Folding the single observations, we have been able to clearly identify it and
confirm its pulsar nature in all the 33 GUPPI observations. As can be seen from
the left panel of Figure~\ref{profili} (see also the top panel of Figure~\ref{sumprof}), Ter5aj presents a double peaked pulse shape,
where the two peaks are separated by $\sim0.3$ in phase. We extracted the pulse times
of arrival (TOAs) with the {$\tt get_{-} TOAs$} routine within PRESTO, using  a 
double Gaussian template, created by fitting the pulses obtained in the
observations where this object has the highest signal to noise ratio (see left panel of Figure~\ref{profili}). {We obtained one TOA per epoch for all the observations but those showing very good detections from which we have been able to obtain multiple TOAs.} We phase connected all the $\sim6$ years of data using standard procedures with TEMPO. {An extra free parameter, called EFAC, has been included to account for underestimations of the TOA uncertainties (and therefore underestimations of the derived parameter uncertainties), thus obtaining a reduced $\chi^2=1.00$.} The timing solution is tabulated in Table~\ref{ajtiming}, the post-fit timing residuals are reported in the top panel of Figure~\ref{resid} {and the averaged pulse profile, obtained by summing all the daily detections, is reported in the top panel of Figure~\ref{sumprof}}.

\begin{figure}
\centering
\includegraphics[width=9cm]{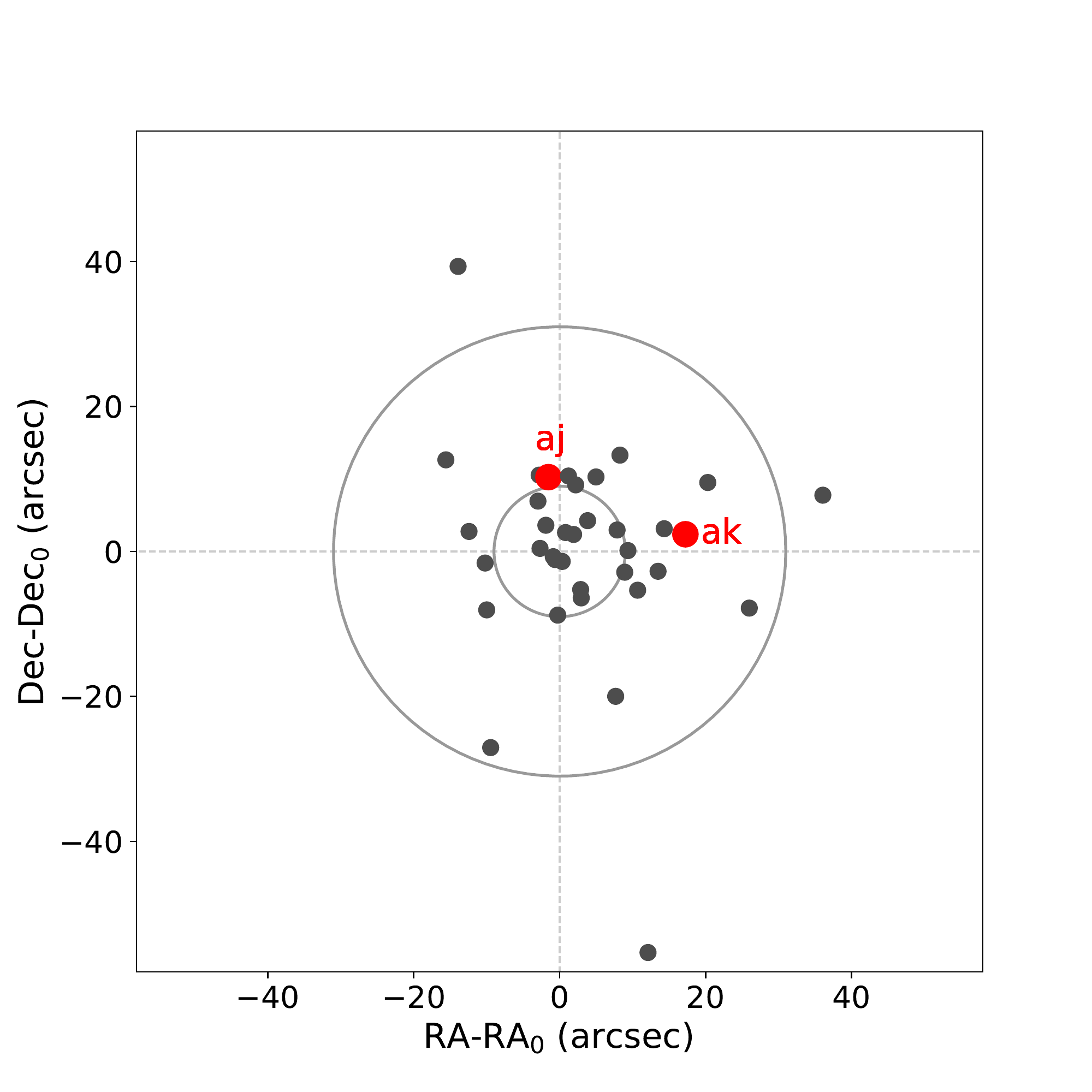}
\caption{Positions of Ter5aj and Ter5ak with respect to the cluster gravitational center and to the other MSPs in the cluster. The inner and outer circles are the cluster core and half mass radius, respectively \citep{lanzoni10}.}
\label{positions}
\end{figure}

Ter5aj is an isolated MSP with a spin period of 2.96 ms and a $\rm DM=238.63 \ pc \
cm^{-3}$, very close to the cluster mean value. It is located $10.4\arcsec$  north
from the cluster gravitational center (\citealt{lanzoni10}, see Figure~\ref{positions}). Its spin period derivative is partially contaminated by the
effect of the MSP motion in the cluster potential field. We will analyze this in more
detail in Section~\ref{intrinsic}. 

%Finally, its proper motion, though poorly constrained, is in agreement with that measured for the other MSPs in the cluster (Ransom et al. 2017, in preparation).

{We roughly estimated the pulsar mean flux density using the radiometer equation {(see Appendix A1.4 of \citealt{lorimer04}) on daily detections, using a system equivalent flux density (SEFD) of 12.5 Jy, appropriate for the observations we used}. The values so obtained have been calibrated by comparison with those obtained applying the same method to other three isolated MSPs (namely Ter5R, Ter5S and Ter5T), for which measurements made referencing a flux calibrator are available (Ransom et al. 2018, in preparation). The average values of both the L-band and S-band flux densities are reported in Table~\ref{ajtiming}. The typical flux density of this object is of the order of that measured for the other faint isolated MSPs of this cluster (Ransom et al. 2018, in preparation)}.

\subsection{Ter5ak}

Ter5ak has been discovered with a maximum stacked spectral power at $\rm DM=236.50
\ pc \ cm^{-3}$. The pulsar nature of this candidate has been confirmed by folding
the single observations, where it turned out to be visible in almost all of them. We managed to obtain a full timing solution for this object { using similar methods described for Ter5aj, except that we obtained the initial timing solution using the}
TEMPO based phase-connection routine available at \url{https://github.com/smearedink/phase-connect}, to be described in Freire \& Ridolfi, in preparation. The timing solution is tabulated in
Table~\ref{ajtiming}, the post-fit residuals are reported in the bottom panel of
Figure~\ref{resid} {and the averaged pulse profile in the middle panel of Figure~\ref{sumprof}}.

Ter5ak is also an isolated MSP and it has a spin period of $\sim1.89$ ms, hence it is
the fourth fastest MSP in Terzan 5 and the fifth fastest among all the GC pulsars. { Its DM of $\rm 236.707 \ pc \ cm^{-3}$ is well within the range covered by the other pulsars in the cluster}. Its position with respect to the other cluster
MSPs is reported in Figure~\ref{positions}, where it can be seen that it is located
at about $17.4\arcsec$ east from the cluster center. As for Ter5aj, its spin period
first derivative is clearly contaminated by its motion in the cluster potential field (see
Section~\ref{intrinsic}).
{The average flux density is also in this case of the order of that of the faintest isolated MSPs of this cluster.}

\subsection{Ter5al}

Ter5al is the last MSP identified in our analysis, with a weak spectral power peaked at
 $\rm DM=236.50 \ pc \ cm^{-3}$. We have been able to reveal this
object in only $\sim20$ observations, being under the detection limit in all the
others. In the right panel of Figure~\ref{profili}, we report the detection plot of
the observation where Ter5al has the highest signal to noise ratio. To date we have
not been able to obtain a timing solution for this system, likely because of the
insufficient number of  good detections. In the $\sim20$ detections we found no
evidence of acceleration, thus it is likely another isolated
system. Ter5al has a spin period of $\sim5.95$ ms. We determined its DM by measuring
the pulse TOAs in different sub-bands of the two brightest {L-band} observations
and we found $\rm DM=~236.48(3) \ pc \ cm^{-3}$.

{The averaged pulse profile is reported in the bottom panel of Figure~\ref{sumprof}. This pulsar turns out to be extremely faint. Indeed its average flux density is of only $\sim8 \mu Jy$ in L-band and $\sim6 \mu Jy$ in S-band, making this pulsar the faintest in the cluster and also explaining the small number of good detections}.

%\vspace{1cm}
\section{Accelerations and physical parameters}
\label{intrinsic}
In this section we derive {some constraints on the accelerations and on the main} physical parameters of Ter5aj and
Ter5ak, the two new MSPs for which we have been able to obtain a timing solution.

For the case of MSPs in GCs, the measured spin period derivative ($\dot P_{meas}$),
derived through timing, does not represent a direct measurement 
of the MSP intrinsic spin-down, since it is the combination of
 different contributions \citep[see, e.g.,][]{phinney93}. Indeed, any motion of
a pulsar with respect to the observer 
%will result in a change of 
produces a change of the
observed spin period. In the case of GCs, the MSP motion in the cluster potential field induces a
change that can be large enough to match or even exceed the value due to the
intrinsic spin-down. Following \citet{phinney93}, $\dot P_{meas}$ can be written as
follows:
\begin{equation}
\label{accels}
\left( \frac{\dot P}{P}\right)_{meas} = \left( \frac{\dot P}{P}\right)_{int} + \frac{a_c}{c} + \frac{a_g}{c} + \frac{a_s}{c}
\end{equation}
where $\left(\dot P/P\right)_{int}$  is the ratio between the intrinsic spin-down and the pulsar
spin period, $a_c$ is the line of sight acceleration due to the GC potential field, $a_g$ is the
acceleration due to the Galactic potential, $a_s$ is an apparent centrifugal acceleration \citep[the
so-called Shklovskii effect;][]{shk70} and $c$ is the speed of light. The two latter terms are expected to be
negligible with respect to the former two, and following \citet[][and references within]{prager17}  
we know that
$a_g= 5.1\times10^{-10} \pm 1.4\times10^{-10} \  \rm m \ s^{-2}$ and $a_s\sim4.2\times10^{-12}  \  \rm m \
s^{-2}$. According to \citet{freire05} and \citet{prager17}, the acceleration along our line of sight ($z$) due to the
cluster potential ($a_c$) can be written as:

%\begin{equation}
\begin{multline}
\label{ac}
a_c (z,x) = -3.5\times10^{-7} \left( \frac{\rho_c}{10^{6} \ M_{\odot} \ pc^{-3}}\right)\left(\frac{z}{0.2 \ pc} \right) \times \\ \times\left(\sinh^{-1}(x) - \frac{x}{\sqrt{1+x^{2}}} \right) x^{-3} \ \rm  m \ s^{-2}
\end{multline}
%\end{equation}
where $\rho_c$ is the cluster core density and $x\equiv r/r^{NS}_c$, where $r^{NS}_c$ is the
cluster core radius of the neutron star population and $r=\sqrt{r_{\perp}^2 +z^2}$ is distance of the pulsar from
the cluster center. In the latter formula, $r_{\perp}=D\theta_{\perp}$ is the pulsar
projected distance from the cluster center, where D is the distance of the cluster
from the Sun and $\theta_{\perp}$ the pulsar angular offset from the cluster center. 

{\citet{prager17} used the ensemble of Terzan~5 MSPs, including Ter5aj and Ter5ak, to
derive the cluster physical properties. They found $\rho_c = 1.58\pm0.13 \times 10^6 \ M_{\odot}
\ pc^{-3}$ and $r_c^{NS} = 0.16\pm0.01$ pc. 
Given the angular offsets of Ter5aj and Ter5ak from the cluster center (see Table~\ref{ajtiming}) and the cluster distance of 5.9
kpc \citep{lanzoni10}, we measured the possible line of sight accelerations of these two MSPs for different values of the line of sight distance $z$.  We found that the maximum allowed accelerations are $\pm 3.4 \times10^{-8} \ \rm m \ s^{-2}$ and $\pm 2.0 \times10^{-8} \ \rm m \ s^{-2}$ for Ter5aj and Ter5ak, respectively. We used these values to constrain, starting from Equation~\ref{accels}, the MSP intrinsic spin-down rates and, consequently, the characteristic ages,
surface magnetic fields and spin-down luminosities. All these values are tabulated in Table~\ref{ajtiming} and, for
both the MSPs, are in agreement with those typically expected for old and recycled pulsars and with that of the other Terzan~5 pulsars.}

\begin{deluxetable*}{lcc}
\tablecolumns{3}
\tablewidth{0pt}
\tablecaption{Timing
parameters for the new Terzan 5 MSPs\tablenotemark{a}.\label{ajtiming}}
\tablehead{\colhead{Parameter} & \colhead{Ter5aj} & \colhead{Ter5ak}}
\startdata
\multicolumn{3}{c}{Timing Parameters} \\ \hline
Right ascension, $\alpha$ (J2000)\dotfill  & 17$^{\rm h}\,48^{\rm m}\,05\fs0119(2)$  & 17$^{\rm h}\,48^{\rm m}\,03\fs6860(2)$    \\
Declination, $\delta$ (J2000)  \dotfill  & $-24^\circ\,46'\,34\farcs85(7)$   & $-24^\circ\,46'\,37\farcs83(8)$  \\
Spin frequency, $F$ (Hz)  \dotfill  &  337.96234149929(4) &  529.07066473956(4) \\
Spin frequency derivative, $\dot F$ $(10^{-14}$ s$^{-2}$) \dotfill & $-$1.61313(7) & $-$2.4771(2)  \\
Spin frequency second derivative, $\ddot F$ $(10^{-25}$  s$^{-3}$)  \dotfill & $-$1.8(4)  & - \\
Dispersion measure, DM (cm$^{-3}$ pc) \dotfill   & 238.633(6)   &  236.705(5)   \\
MJD range \dotfill  & 55423-57573 &  55423-57573  \\ 
Epoch (MJD) \dotfill & 56498 & 56498  \\
Data span (yr) \dotfill  & 5.9 &  5.9  \\
Number of TOAs \dotfill & 42  & 31   \\
RMS timing residuals ($\mu s$) \dotfill  & 12.51 & 17.03  \\ 
EFAC \dotfill & 1.14 & 1.18 \\
Reduced $\chi^{2}$ value \dotfill & 1.00 & 1.00 \\
Solar system ephemeris model \dotfill & DE436 & DE436 \\
\hline
\\
\multicolumn{3}{c}{Derived Parameters} \\ \hline
Spin period, $P$ (ms)  \dotfill  & 2.9589095505841(3) &   1.8901066845055(1) \\
Spin period derivative, $\dot P$ ($10^{-19}$) \dotfill & 1.41232(6) & 0.88495(6)  \\ 
Spin period second derivative, $\ddot P$ ($10^{-30}$ s$^{-1}$) \dotfill & 1.6(3) & - \\ 
Intrinsic spin period derivative, $\dot P_{\rm int}$ $(10^{-19})$\dotfill & $ <4.9$ &  $<2.1$\\
Characteristic age, $\tau_c$ (Myr) \dotfill & $>98$ & $>142$ \\
Surface magnetic field, $B_0$ ($10^8$ G) \dotfill &  $ <12.0$  & $<6.4$\\
Spin-down luminosity, $L_{SD}$ ($10^{35}$ erg s$^{-1}$) \dotfill & $<7.3$ & $<12.4$ \\
Flux density at 1.5 GHz, $S_{1.5}$ ($\mu Jy$)  \dotfill & 34 & 30 \\
Flux density at 2.0 GHz, $S_{2.0}$ ($\mu Jy$)  \dotfill & 18 & 16 \\
Angular offset from cluster centre, $\theta_{\perp}$ (\arcsec) \dotfill & 10.4 & 17.4 \\
\hline
\enddata
\tablenotetext{a}{Numbers in parentheses are uncertainties in the last digits
quoted. The time units are TDB and the adopted terrestrial time standard is UTC(NIST).}
\end{deluxetable*}

\section{SUMMARY AND CONCLUSIONS}
\label{conclusions}

We used archival observations to search for new pulsars in the stellar system
Terzan 5. Instead of using classical search routines based on the analysis of
single observations, we developed an alternative method which combines multiple observations. In this method, 
 stacked power spectra at different DMs are created by summing the spectral
powers obtained in the single observations.  In order to remove RFI, a control DM spectrum has been subtracted from all the stacked power spectra.
% have been  obtained
%in a range of DMs compatible with the cluster one and they have been subtracted by a
%stacked power spectrum obtained at a control DM (i.e., about half the cluster mean
%value) to remove a large fraction of RFI. 
All the candidates selected in these ``corrected" stacked power spectra have been
processed with a KD-Tree algorithm, in order to select those that are likely
pulsar candidates and discard those that are more likely persistent RFI. 
This method turns out to be a quite powerful tool to search for very faint isolated
pulsars in GCs and it opens the possibility to identify  a significant
number of still unknown pulsars by simply using the large amount of archived data
collected in the last decades of GC observations. Together with the classical routine
searches on single observations, it could be a complementary method to discover
extremely faint pulsars by using the large amount of GC data that the new generation
of radio telescopes (such as MeerKAT) are going to produce.

{ The stacking technique described in this work is of particular effectiveness in a high DM system like Terzan 5. Indeed, for typical 1.4 GHz observations, the high DM and very small scintillation bandwidth allow us to average, in the stacked power spectra, over many scintles, making negligible any effect due to diffractive scintillation. The only variability appreciable in the different terms of the power spectra sum is due to refractive scintillation, which can affect the flux densities of pulsars by typically up to a factor of $\sim2$. On the other hand, in a low DM cluster such as, for example, 47 Tucanae, diffractive scintillation can change the measured flux densities by more than an order of magnitude. Therefore few of the power spectrum sums will have high values, while most of them will have much lower values, thus diluting the signal to noise ratio of the final sum. However, even
in such conditions stacked power spectrum searches
can reveal new pulsars, as demonstrated by \citet{pan16}.}

The application of the method to Terzan 5
 led us to discover three additional MSPs in this stellar system.
For two MSPs, we have been able to obtain a phase connected timing solution that confirmed their
association with the GC, having a DM close to the cluster mean value and being
located within $\sim17\arcsec$ from the cluster gravitational center. 
These discoveries  bring the total
number of known MSPs in this system to 37, $\sim 25\%$ of the entire
pulsar population identified so far in GCs.  

Indeed Terzan~5 turns out to be the most efficient factory of MSPs in
the Milky Way and the large number of  X-ray sources (see, e.g.,
\citealp{heinke06}) and  X-ray bursters (see the recent case of 
EXO 1745-248, \citealt{altamirano15,ferraro15}) suggest that this furnace is
currently quite active. 
 
\citet{ferraro09} first pointed out that, at odds with what is
commonly thought, Terzan 5 probably is not a genuine GC, 
but a much more complex stellar system, since it hosts different stellar 
populations characterized by significantly different
iron abundances (see also \citealp{origlia11, origlia13, massari14}). 
Recently \citet{ferraro16} measured the ages of the two main 
sub-populations, (finding 12 and 4.5 Gyr for the sub-solar and super-solar metallicity
component, respectively), thus identifying  Terzan 5 as a site of
multiple bursts of star formation in the Galactic bulge.
%\footnote{ Note that these results definitely  confirmed that Terzan 5 is not a genuine
%CGs. Indeed, apart from the overall morphological appearance, Terzan 5
%does not share any other characteristics of globulars, neither in terms of chemistry
%(GCs show inhomogeneities only in the light-elements;
%\citealt{carretta09}), nor in the enrichment history and age spread of
%the sub-populations (the enrichment timescale in GCs is of a few
%$10^8$ yrs and their light-element sub-populations are thus almost
%coeval; \citealt{dercole08}), nor in terms of the mass of the
%progenitor (GCs did not retain the high-velocity SN ejecta and
%therefore do not require particularly massive progenitors). }.
 In fact, 
the measured chemical patterns
and the large age difference between 
the two main sub-populations could be naturally explained 
in a self-enrichment scenario where
Terzan 5 was originally much more massive ($\sim 10^8
M_\odot$) than today ($\sim 10^6
M_\odot$; \citealp{lanzoni10}), and therefore able to retain the
iron-enriched gas ejected by violent supernova explosions.  The large
number of type II supernovae required to explain the observed
abundance patterns should have also produced a large population of
neutron stars, mostly retained into the deep potential well of the
massive {\it proto}-Terzan 5 and likely forming binary systems through
tidal capture interactions.  This, together with its large collision rate
\citep{lanzoni10}, the largest among all Galactic globular clusters
\citep[see also][]{verhut87}, could have highly promoted pulsar
re-cycling processes, which explains the production of the large
population of MSPs and other stellar exotica now observed in the system.

\section*{ACKNOWLEDGEMENT}
M.C. thanks the
{\it Marco Polo Project}
of the Bologna University
and the National Radio Astronomy Observatory for the hospitality during his stay in
Charlottesville, where part of this work was carried out. \\
The National Radio Astronomy Observatory and the Green Bank Observatory are facilities of the National Science Foundation operated under cooperative agreement by Associated Universities, Inc.  S.M.R. is a fellow of the Canadian Institute for Advanced Research and receives additional support from NSF Physics Frontier Center award number 1430284.\\
J.W.T.H. acknowledges funding from an NWO Vidi fellowship and from
the European Research Council under the European Union's Seventh
Framework Programme (FP/2007-2013) / ERC Starting Grant agreement
nr. 337062 ("DRAGNET").\\
Pulsar research at UBC is supported by an NSERC Discovery Grant and by the Canadian Institute for Advanced Research.\\

\facility{GBT}

\software{Scipy \citep{jones01}, PRESTO \citep{ransom02}, TEMPO \citep{manchester15}}

\end{document}